\documentclass[conference,10pt]{IEEEtran}
\IEEEoverridecommandlockouts
\usepackage{cite}
\usepackage{amsmath,amssymb,amsfonts}
\usepackage{algorithmic}
\usepackage{graphicx}
\usepackage{textcomp}
\usepackage[dvipsnames]{xcolor}
\usepackage{xspace}
\def\BibTeX{{\rm B\kern-.05em{\sc i\kern-.025em b}\kern-.08em
    T\kern-.1667em\lower.7ex\hbox{E}\kern-.125emX}}

\usepackage{verbatim} 
\usepackage{balance}
\usepackage{soul}

\definecolor{olivegreen}{rgb}{0, 0.6, 0}

\newcommand{\JL}[1]{{\color{magenta}[\textbf{\sc JLee}: \textit{#1}]}}
\renewcommand{\JL}[1]{}
\newcommand{\aligner}{SALoBa\xspace}

\newcommand{\sect}[1]{Section~\ref{sec:#1}} 
\newcommand{\fig}[1]{\figurename~\ref{fig:#1}}
\newcommand{\tab}[1]{\tablename~\ref{tab:#1}}
\newcommand{\eq}[1]{Eq.~\ref{eq:#1}}

\usepackage{textcomp}
\usepackage{subcaption}
\usepackage{booktabs}

\usepackage{fancyhdr}

\usepackage{marginnote}
\usepackage[bookmarks=true,breaklinks=true,letterpaper=true,colorlinks,linkcolor=black,citecolor=blue,urlcolor=black]{hyperref}

\newcommand{\revref}[1]{\hyperref[rev:#1]{\color{magenta}#1}}
\newcommand{\figref}[1]{\hyperref[fig:#1]{\color{blue}\figurename~\ref{fig:#1}}}
\newcommand{\tabref}[1]{\hyperref[tbl:#1]{\color{blue}\tablename~\ref{tab:#1}}}

\usepackage{pgfplots}
\usepackage{pgfplotstable}
\usepackage{tikz}
\usetikzlibrary{backgrounds}
\usetikzlibrary{positioning}
\usetikzlibrary{matrix}
\usetikzlibrary{shapes}
\usetikzlibrary{plotmarks}

\pgfplotsset{
    discard if/.style 2 args={
        x filter/.code={
            \edef\tempa{\thisrow{#1}}
            \edef\tempb{#2}
            \ifx\tempa\tempb
                
            \fi
        }
    },
    discard if not/.style 2 args={
        x filter/.code={
            \edef\tempa{\thisrow{#1}}
            \edef\tempb{#2}
            \ifx\tempa\tempb
            \else
                
            \fi
        }
    }
}
\newcounter{groupcount}
\pgfplotsset{
    draw group line/.style n args={5}{
        after end axis/.append code={
            \setcounter{groupcount}{0}
            \pgfplotstableforeachcolumnelement{#1}\of\datatable\as\cell{%
                \def\temp{#2}
                \ifx\temp\cell
                    \ifnum\thegroupcount=0
                        \stepcounter{groupcount}
                        \pgfplotstablegetelem{\pgfplotstablerow}{[index]0}\of\datatable
                        \coordinate [yshift=#4] (startgroup) at (axis cs:\pgfplotsretval,0);
                    \else
                        \pgfplotstablegetelem{\pgfplotstablerow}{[index]0}\of\datatable
                        \coordinate [yshift=#4] (endgroup) at (axis cs:\pgfplotsretval,0);
                    \fi
                \else
                    \ifnum\thegroupcount=1
                        \setcounter{groupcount}{0}
                        \draw [
                            shorten >=-#5,
                            shorten <=-#5
                        ] (startgroup) -- node [anchor=north] {\footnotesize{#3}} (endgroup);
                    \fi
                \fi
            }
            \ifnum\thegroupcount=1
                        \setcounter{groupcount}{0}
                        \draw [
                            shorten >=-#5,
                            shorten <=-#5
                        ] (startgroup) -- node [anchor=north] {\footnotesize{#3}} (endgroup);
            \fi
        }
    }
}

\usepgfplotslibrary{groupplots}

\definecolor{black}{HTML}{000000}
\definecolor{white}{HTML}{ffffff}
\definecolor{color1}{HTML}{90ee90}
\definecolor{color2}{HTML}{F0E68C}
\definecolor{color3}{HTML}{DCD0FF}
\definecolor{color4}{HTML}{B19CD9}
\definecolor{color5}{HTML}{FFB6C1}
\definecolor{color6}{HTML}{20B2AA}
\definecolor{color7}{HTML}{87CEEB}
\definecolor{color8}{HTML}{FFA07A}

\definecolor{gray0}{rgb}{0.95, 0.95, 0.95}
\definecolor{gray1}{rgb}{0.1, 0.1, 0.1}
\definecolor{gray2}{rgb}{0.3, 0.3, 0.3}
\definecolor{gray3}{rgb}{0.5, 0.5, 0.5}
\definecolor{gray4}{rgb}{0.7, 0.7, 0.7}
\definecolor{gray5}{rgb}{0.9, 0.9, 0.9}

\pgfplotsset{compat=newest}
\makeatletter
\newcommand\resetstackedplots[1]{
\makeatletter
\pgfplots@stacked@isfirstplottrue
\makeatother
\pgfplotstablenew[
  create on use/x/.style={create col/expr={\pgfplotstablerow}},
  create on use/y/.style={create col/expr={0}},
  columns={x,y}]{#1}\zerotable
\addplot [forget plot,draw=none] table[x=x,y=y] from \zerotable;
}
\makeatother

\begin{document}





\pagenumbering{arabic}
\thispagestyle{plain}
\pagestyle{plain}
\setcounter{page}{1}

\title{\aligner: Maximizing Data Locality and Workload Balance for Fast Sequence Alignment on GPUs }

\author{%
\IEEEauthorblockN{%
Seongyeon Park\IEEEauthorrefmark{2},
Hajin Kim\IEEEauthorrefmark{2},
Tanveer Ahmad\IEEEauthorrefmark{4},
Nauman Ahmed\IEEEauthorrefmark{4},\\%
Zaid Al-Ars\IEEEauthorrefmark{4},
H. Peter Hofstee\IEEEauthorrefmark{4}\IEEEauthorrefmark{3},
Youngsok Kim\IEEEauthorrefmark{2},
and Jinho Lee\thanks{\textsuperscript{*}Corresponding author}\IEEEauthorrefmark{2}\textsuperscript{*}} 
\IEEEauthorblockA{%
\IEEEauthorrefmark{2}\textit{Yonsei University}\hspace{15pt}%
\IEEEauthorrefmark{4}\textit{TU Delft}\hspace{15pt}%
\IEEEauthorrefmark{3}\textit{IBM}}%
\{syeonp, kimhajin\},@yonsei.ac.kr, \{t.ahmad, n.ahmed, z.al-ars\}@tudelft.nl,\\%
hofstee@us.ibm.com, \{youngsok, leejinho\}@yonsei.ac.kr}


\maketitle

\begin{abstract}
Sequence alignment forms an important backbone in many sequencing applications. 
A commonly used strategy for sequence alignment is an approximate string matching with a two-dimensional dynamic programming approach.
Although some prior work has been conducted on GPU acceleration of a sequence alignment, we identify several shortcomings that limit exploiting the full computational capability of modern GPUs.
This paper presents \aligner, a GPU-accelerated sequence alignment library focused on seed extension.
Based on the analysis of previous work with real-world sequencing data, we propose techniques to exploit the data locality and improve workload balancing.
The experimental results reveal that \aligner significantly improves the seed extension kernel compared to state-of-the-art GPU-based methods.
\end{abstract}


\section{Introduction}


With the fast advances in next-generation sequencing (NGS) techniques, the monetary cost for DNA sequencing has been declining at a rate that is outpacing Moore's law~\cite{dnacost}.
However, on the other side of the coin, this rapid throughput in sequencing means that data processing has become a more severe bottleneck.
For example, performing read mapping of the human genome on an Intel Xeon processor now takes more than 20$\times$ the sequencing time~\cite{goyal2017ultra}.

\emph{Read mapping}, a process in sequence alignment, maps a piece of query DNA (generated from sequencing) to matching locations of a reference DNA.
However, the reference and query DNAs do not exactly match due to sequencing errors and/or mutations.
Therefore, the problem falls into the category of approximate string matching.

One popular strategy to solve this problem is \emph{seed-and-extend}, where the seeding phase locates a few exact matches, and the extension phase performs approximate matching based on dynamic programming (DP) such as the Smith--Waterman algorithm~\cite{sw}.
Unfortunately, the Smith--Waterman algorithm has quadratic complexity with the input length, which continues to increase as NGS techniques evolve.

In this work, we focus on the extension, which consumes a significant portion of the execution time for read mapping~\cite{primer}. 
Many proposals have been made for dedicated hardware accelerators with ASICs~\cite{fujiki2018genax,turakhia2018darwin,fujiki2020seedex} or FPGAs\cite{ham2020genesis, houtgast2015fpga,zhang2007implementation,ahmed2015heterogeneous,blastp} to cope with the computational bottleneck problem.
However, such accelerators are yet to be widely used in practice. 
The ASICs are expensive to produce and would have difficulty appearing in the market unless mass production was guaranteed.
The FPGAs are less prone to this issue because many FPGA accelerator cards are already available on the market. 
However, FPGA accelerator cards are not yet the dominant option because they are expensive and hard to program. 

Under such circumstances, GPU-based accelerations can be viable options because they are cheaper, easier to find on the market, and require less effort to develop software. 
Therefore, many GPU-based libraries have been developed for bioinformatics~\cite{cushaw1,gasal2,nvbio}. 
With the fast growth in the computational capabilities of GPUs~\cite{ampere}, GPUs will likely continue to be the primary option for accelerating sequence alignments.

However, by analyzing the state-of-the-art GPU-accelerated seed extension, we identified several missing opportunities for performance improvement that have become critical, especially with long string queries.
Considering that the lengths of sequence reads are rapidly growing with third-generation sequencers~\cite{third}, the performance gap between the ideal case and the existing software will widen. 
Specifically, we found that existing GPU kernels 1) inefficiently utilize memory and 2) suffer from load imbalance.  

In this paper, we present \aligner (\underline{\textbf{S}}equence \underline{\textbf{A}}lignment with Data \underline{\textbf{Lo}}cality and Workload  \underline{\textbf{Ba}}lance), which addresses the mentioned problems to achieve superior performance. 
Taking lessons from several ASIC/FPGA accelerators~\cite{ahmed2015heterogeneous,zhang2007implementation, fujiki2020seedex}, \aligner puts together a kernel that makes better use of the computational capability of modern GPUs.

\aligner utilizes intra-query parallelism by allocating multiple CUDA threads to a query-reference pair.
Even though this option has been studied by some prior art~\cite{korpar2013sw,adept}, 
these suffer from inefficient memory access patterns and resource underutilization, being outperformed by the current state-of-the-art, which only uses inter-query parallelism.
\aligner addresses those issues by introducing two novel techniques: lazy spilling to global memory and subwarp scheduling.
In \emph{lazy spilling}, the data are first accumulated to the CUDA shared memory and later spilled to the global memory in a coalesced manner. 
By utilizing the double-buffered shared memory region in a rotating manner, the redundancy in global memory is greatly reduced without much overhead to the shared memory.
In addition, \emph{subwarp scheduling} divides a warp into multiple subwarps to mitigate the underutilization problem.
While this slightly increases the workload imbalance, the gain from better resource utilization often dominates the overhead from workload imbalance.

According to the experimental results, \aligner performs significantly faster than the state-of-the-art GPU aligner in the seed extension kernel over several sequence lengths.
When tested on real-world datasets with various lengths, we obtained superior results due to better workload balancing.
Our contributions can be summarized as follows:
\begin{itemize}
    \item We identify several unexploited opportunities from the current libraries for performance improvements.
    \item We propose \aligner, which provides significant speedup compared to the current state-of-the-art GPU-based alignment libraries.
    \item We perform an extensive amount of evaluation, including synthetic and real-world data from a popular read alignment software to demonstrate the efficiency of \aligner.
\end{itemize}


\section{Background}

\subsection{Seed Extension}
The heart of the sequence alignment problem is approximate string matching. Because there could be multiple possible paths to take every time a mismatch occurs, its complexity quickly increases along with the length of the input pair. 

The seed-and-extend strategy is an approach to perform alignment efficiently. Based on the observation that a good alignment usually contains many matches, the strategy first finds multiple exact matches, called \emph{seeds}. 
Based on these, approximate matching scores are calculated by \emph{extending} to both directions from the found seeds.

\begin{figure}
    \centering
    \includegraphics[width=0.95\columnwidth]{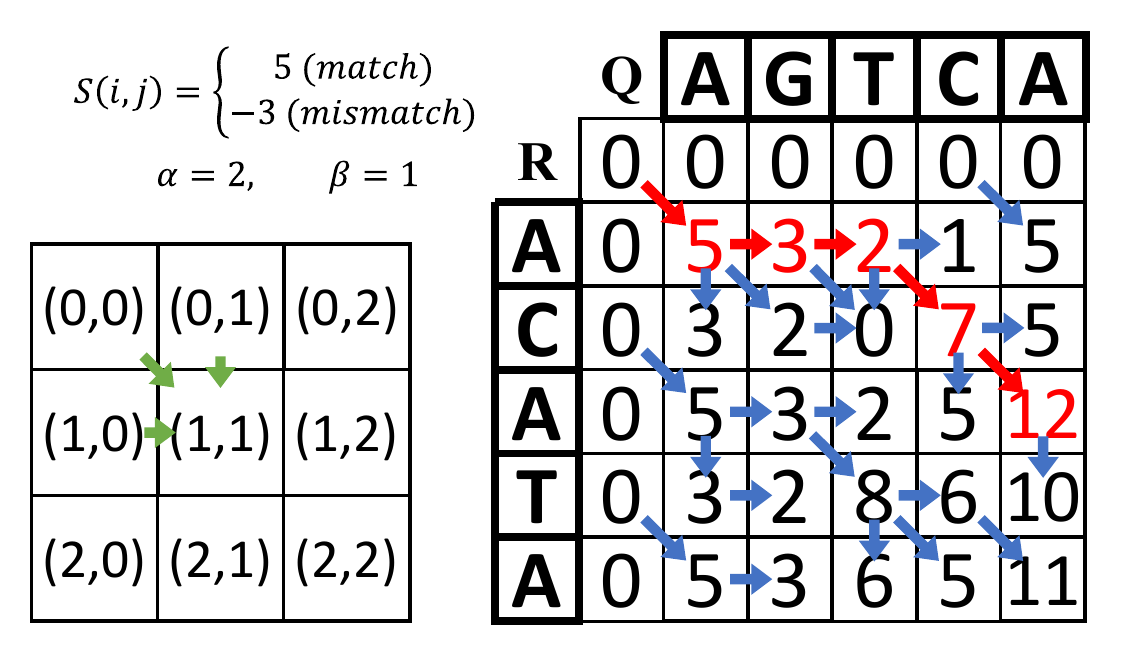}
    \caption{\vspace{-4mm}An example seed extension algorithm.\vspace{-1mm}}
    \label{fig:sw}
\end{figure}

Seed extension is often performed using a DP algorithm that fills a two-dimensional DP table with the complexity of $O(N^2)$. 
Popular algorithms are Smith--Waterman~\cite{sw} and Needleman--Wunsch~\cite{nw}. 
The score of each cell is calculated based on predetermined scores for the type of differences (i.e., insert, delete, and mismatch).  
With some adjustments for considerations for frequent long gaps, the affine gap function calculates the score of a DP cell $H(i, j)$ as follows:
\begin{align}
    H(i,j) &= max\begin{cases}0\\
    E(i,j)\\
    F(i,j)\\
    H(i-1,j-1) + S(i,j)    
    \end{cases},\\
    E(i,j) &= max\begin{cases}    H(i,j-1) - \alpha \\
    E(i,j-1)-\beta \\
    \end{cases},\\
    F(i,j) &= max\begin{cases}    H(i-1,j) - \alpha \\
    F(i-1,j)-\beta \\
    \end{cases},
    \label{eq:sm}
\end{align}
where $S(i, j)$ is a score function that returns a positive value when the reference at $i$ and query at $j$ match and returns a negative value otherwise. 
In addition, $E$ and $F$  are auxiliary variables that keep track of the continuing gaps. 
Last, $\alpha$ and $\beta$ account for different gap penalties according to new gaps or continued gaps, respectively.

\fig{sw} presents an example DP table, where $Q$ represents a query string, and $R$ represents a reference string. 
To calculate a cell, the cells from three adjacent cells, the top, left, and top left, are needed, represented by green arrows. 
The trace of the highest score in the DP table represents the best match found by the algorithm, denoted with red arrows and numbers. 

\subsection{Baseline GPU-based Seed Extension}
\label{sec:gasal}

There have been several attempts to accelerate seed extension using GPUs.
\JL{change this to intra-query and inter-query parallelism. Also talk about 2bit issues.}
In sequence read data, each base is represented by a character data type of eight bits.
However, only five bases exist within the sequences: A, C, G, T/U, and N, where T is used for DNA and U is used for RNA. 
The N denotes unknown bases. 
Having five bases indicates that at least three bits are needed to represent each.
Because three-bit representations are inefficient to deal with in modern architectures, a four-bit packed representation is often used~\cite{gasal2,nvbio} and some work utilizes eight-bit representation~\cite{korpar2013sw}.

Existing methods for GPU-based seed extension can be roughly categorized by the parallelism they utilize: \emph{intra-query parallelism} and \emph{inter-query parallelism}.
Intra-query parallelism refers to processing multiple cells in a DP table concurrently. 
As shown in \fig{sw}, parallelism exists in an anti-diagonal form in the DP table. For example, after the cell (0,0) has been processed, cells (0,1) and (1,0) can be processed in parallel.
Afterwards, cells (0,2), (1,1), and (2,0) can be processed in parallel. 
Many ASIC/FPGA accelerators successfully take advantage of this, commonly using one-dimensional systolic array structures~\cite{zhang2007implementation,fujiki2020seedex,ahmed2015heterogeneous}.
Some GPU-based approaches utilize this type of parallelism~\cite{korpar2013sw,adept}.
However, they often fail to achieve sufficient speedup due to resource under-utilization and inefficient memory access~\cite{adept}.
On the other hand, inter-query parallelism refers to simply processing multiple query-reference pairs in parallel.
Because the latter is easier to optimize resource utilization, many successful libraries adopt this approach~\cite{gasal2,nvbio,cushaw2gpu,soap3}.


Among the libraries using inter-query parallelism, GASAL2~\cite{gasal2} is known to show the current state-of-the-art performance.
In this section, we describe the details of its strategy.
The GPU registers are 32-bit wide; thus, eight bases from each query and reference are fetched in a single step.  
Therefore, it is natural to process 8\texttimes8 cells at once that correspond to a single-word reference and a single-word query. 
After processing the block of 8\texttimes8 cells, the thread advances to the right. 
For this, a column of eight rightmost cells must be stored for dependency. The dependency is fulfilled by keeping the values of these DP cells within the registers. 
When a thread reaches the rightmost end of the table, it moves to the bottom part of the DP table. 
Then, the cells from the top must be accessed for dependency, and the thread stores the bottom eight cells of each 8\texttimes8 block in the global memory. 
Thus, considering the matrix size of $N\times N$, the amount of global memory access becomes $N\times N/4$ for reading and writing dependent cell values.

\input{eval/histo250}

\begin{figure*}
    \centering
    \includegraphics[width=\textwidth]{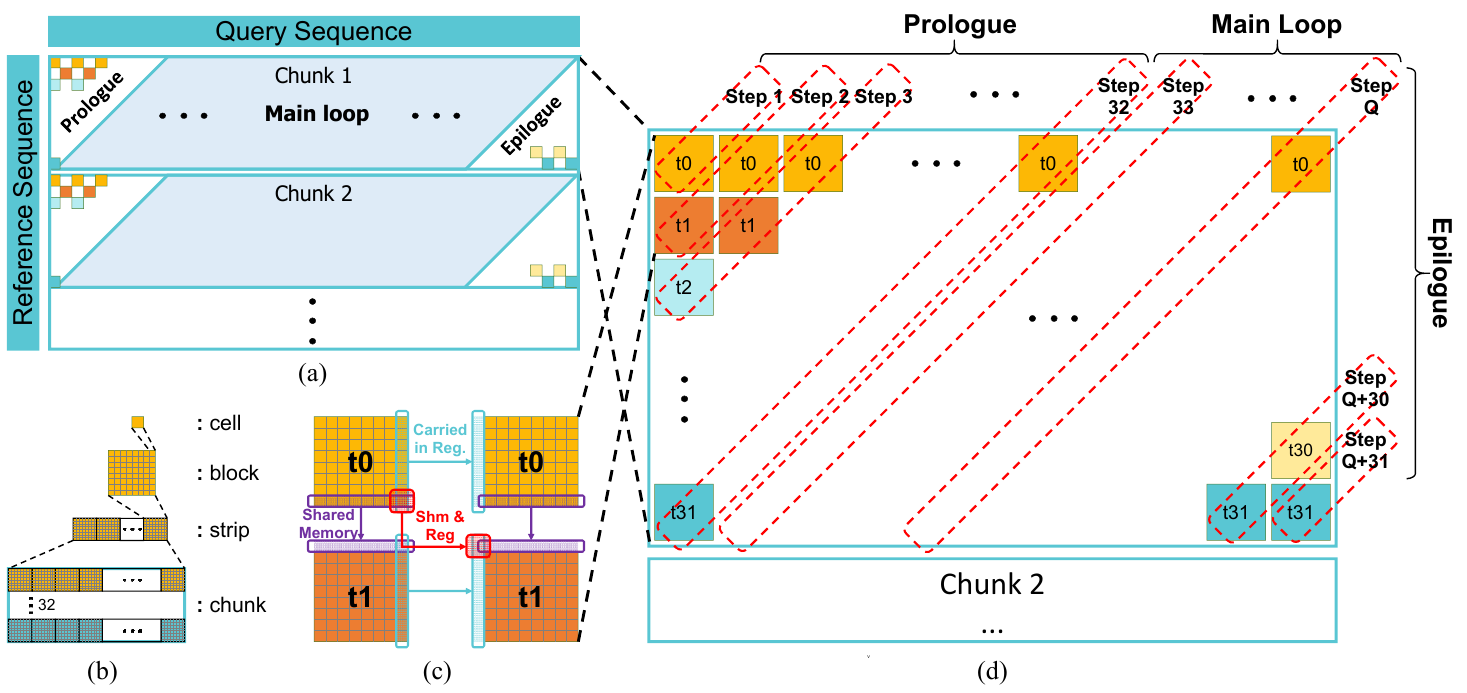}
    \caption{Intra-query parallelism. (a) shows parallelization strategy using a warp, (b) illustrates the terms used in this figure. (c) demonstrates how the dependency between cells are handled and (d) depicts the use of anti-diagonal structure.}
    \label{fig:parallel}
\end{figure*}

\section{Motivation - Diagnosis}
\label{sec:moti}
In this section, we diagnose the state-of-the-art aligner GASAL2~\cite{gasal2} and discuss potential opportunities for further performance improvement. 
First, we reveal a significant load imbalance between individual queries. 
Second, we identify that the use of global memory is not coalesced, which requires a significant amount of redundant memory access.

\subsection{Load Imbalance}
As discussed in Section~\ref{sec:gasal}, the main parallelization scheme of GASAL2 is inter-query parallelism.
In the strategy of letting each CUDA thread handle a single query, the main advantage is that there is no complicated inter-thread synchronization or costly communication.

However, one drawback of this approach is that it ignores the variance in the workload for the seed extension. 
The seeding step provides the query and reference sizes as input to the seed extension. 
Because of this, the lengths vary by individual inputs, and a significant imbalance occurs.

\fig{histo250} plots the distribution of the query and reference sequences from two types of reads (see \sect{eval} for detailed settings) from the popular alignment software BWA-MEM~\cite{bwamem}. 
As illustrated, both distributions range from zero to several hundred or thousand and are not well clustered, implying a substantial amount of warp divergence within GPUs.  
As depicted in the figure, the difference of length between the shortest and longest strings can be up to $10\times$ for both the query and reference string. 
Provided that the computational complexity is proportional to the multiple of the two string lengths, the workload imbalance could be very large in practice.

\subsection{Memory Inefficiency}
With four-bit sequence packing, it is rational to process $8\times8$ DP cells at a time. 
When moving to the next cell,\footnote{Without loss of generality, we assume that cells are processed left to right and top to bottom.} the current cell content can be captured within the registers. 
However, the dependency structure of the seed extension also requires long-term storage of the cell information to process the next row of $8\times8$ cells. 
This intermediate data well exceed the size of the register file and are  usually stored in the global memory (i.e., DRAM).

However, this scheme incurs two inefficiencies. 
First, the intermediate data need not remain in the global memory at the end of the kernel and are considered overhead.  
Therefore, reducing this overhead contributes to better performance. 
Second, the minimum access size of the GPU’s global memory is 128 B (or 32 B from Volta~\cite{volta} architecture, as identified by~\cite{khairy2018exploring}), whereas the individual cell data size is only 4 B. 
If not captured by the L2 cache, this will incur a number of redundant access instances for the same intermediate data, leading to an inefficient kernel.

\begin{table}[t]
\normalsize
    \centering
    \caption{Amount of Data Stored and Accessed for the Existing GPU Aligner}
    \begin{tabular}{lc}
    \toprule
        Data & Quantity \\
        \midrule
        Necessary & $2N$ \\
        Stored & $2N + N^2/4 $ \\
        Accessed (Until Pascal~\cite{pascal}) &$128N + 16N^2$  \\
        Accessed (After Volta~\cite{volta}) & $32N + 4N^2$  \\
        \bottomrule
    \end{tabular}
    \label{tab:moti}
\end{table}

\tab{moti} lists  the data volume, input data necessary for the extension (Necessary), data stored for intermediate data (Stored), and data accessed due to the access granularity (Accessed). 
The table reveals that the inefficiency is multifold even with modern architectures, which provides another opportunity for improvements. 
In this work, we demonstrate that \aligner can remove much of this access, leading to superior performance.




\section{\aligner Design}
\subsection{Intra-query Parallelism}
\label{sec:parallel}
Because each cell in the DP table has dependencies from the top, left, and top-left cells, it is known that intra-query parallelism exists in an anti-diagonal form, as depicted in \fig{parallel}. 
\JL{chk the sentence here}
Adopting the idea, we use multiple CUDA threads to concurrently handle anti-diagonal elements. 
A slight difference is that instead of individual threads computing a single cell at a time, the cells are assigned to threads in $8\times8$ \emph{blocks} because of the 4-bit packing on the reference and query sequences.

In the first version of the library, we decided that 32 threads at most---a warp---should collaborate in processing a query. 
Because intra-warp synchronization is relatively cheap or free (for GPUs before introduction of independent thread scheduling~\cite{volta}), this becomes an attractive design choice.
While using more than 32 threads is theoretically possible, it would require threadblock level synchronizations (i.e., \texttt{\small\_\_syncthreads()}) that cause non-negligible overhead.


For communication between the threads, we used the CUDA shared memory. 
Using shared memory as a communication channel and reusing it every iteration, only a fixed amount of shared memory is needed, and all access to the shared memory is conflict-free (see \sect{shm}).

\fig{parallel} (a) and (b) illustrate the procedure of \aligner.  
The largest units of computation are called \emph{chunks},  which are horizontal partitions of the DP table. 
The chunks have the height of 32 blocks and the width of the entire query sequence.

\begin{figure*}[t]
    \includegraphics[width=0.96\textwidth]{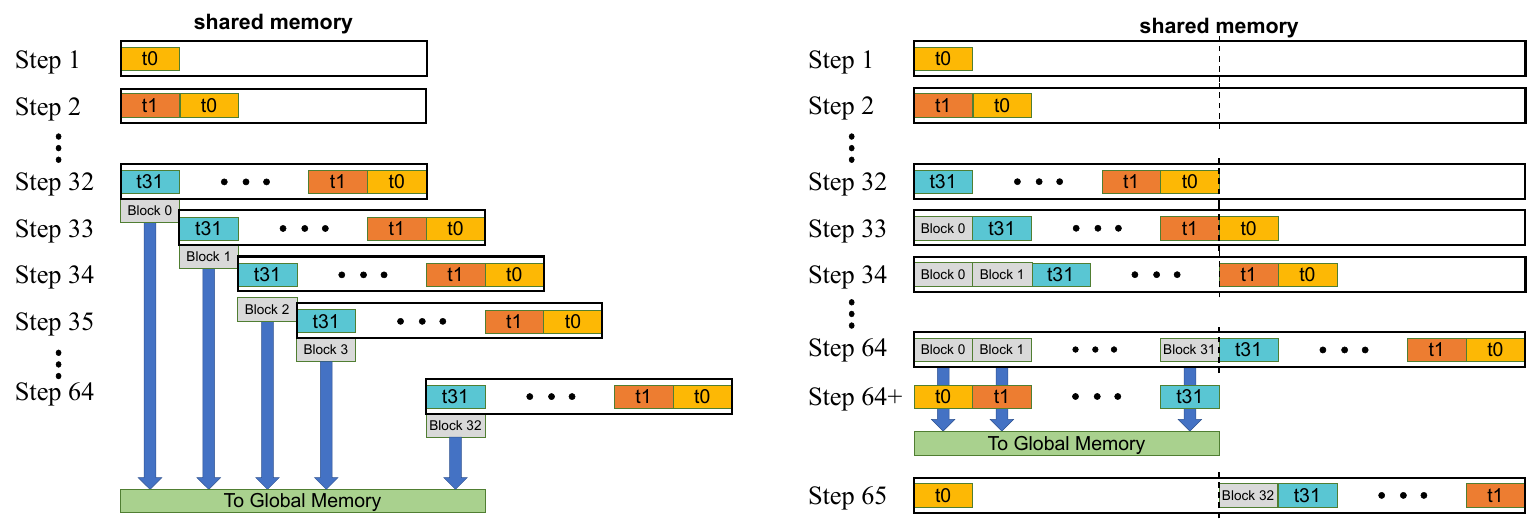}
    \caption{Lazy spill optimization. (Left): Naive global memory access. (Right): Proposed optimized access. Each small rectangle represents 8\texttimes1 cells stored for dependency. A colored rectangle labeled `t\#' denotes that it is being accessed by thread\# and a gray rectangle labeled `block \#' represent that it is from block id \# and is ready to be spilled to the global memory.} 
    \label{fig:shm}
\end{figure*}

As displayed in the figure, a thread processes a block in a \emph{strip} (i.e., a row of blocks) of the chunk per step. 
In the first step, the first thread starts the processing by fetching a 32-bit word from each of the query and reference sequences, which is enough to process an  $8\times$8 \emph{block} of cells.

The number of ready-to-process cells increases by one per step in the upcoming steps.
Therefore, the number of threads participating in the computation increases until the 32nd step, forming a 31-step long \emph{prologue}.
In the main loop stage,\footnote{This pattern is also called the `kernel,’ but we decided to call it the main loop stage because it would be confusing with the term GPU kernel.} all threads in the warp simultaneously process blocks in an anti-diagonal manner, advancing one block to the right per step. 
This process  continues for $Q - 31$ steps, where $Q$ is the number of the blocks in the query sequence (i.e., $\lceil query\_length/8\rceil$).
When the first thread reaches the end of the row, the \emph{epilogue} starts, and the number of participating threads decreases by one per step until the last thread reaches the end of the row. 
At the end of the epilogue, the total number of steps taken for processing the entire chunk is $Q + 31$.

Each time the threads advance to the next step, they access the dependency data from the cells on the top, left, and top-left (\eq{sm}). 
\fig{parallel} (c) demonstrates how this is done.
When a block is calculated, the data from the eight cells to the left are what the current thread computed in the previous step. 
Therefore, the data can be stored in the register. 
%
The data from the eight cells at the top were computed by the adjacent thread in the previous step. 
At the end of processing a block, each thread writes the bottom 8\texttimes1 cells in the shared memory.
In the next step, each thread reads the data in the next position of the shared memory so that it can receive the 8\texttimes1 cells from the above block. \JL{this paragraph has been edited a lot by the editor. Check for its meaning again}
Last, one cell at the top left has dependency on the current block, which is what the thread received from the shared memory in the previous step. 
Fortunately, it can also be passed using the register. 
Thus, the number of cells stored in the register becomes nine instead of eight.

This scheme not only reduces the warp divergence but also improves memory access. 
In the previous technique, all bottom cells in each strip must be stored and read back to/from the global memory. 
In contrast, with intra-query parallelism, only the bottom cells of each chunk (not of the strip) are stored and read to/from the global memory. 
For a 32-thread warp, this reduces the amount of intermediate data access to 1/32.

\subsection{Lazy Spill to Global Memory Access}
\label{sec:shm}
For processing a chunk, the data from the bottom-most cells must be stored for processing the top-most cells of the next chunk. 
Usually, this easily exceeds the shared memory capacity and must be stored in the global memory. 
In a naive scheme, as in \fig{shm} (left), the last thread ($t31$) stores the bottom-most cells of a block in the global memory, and the first thread ($t0$) reads them from the global memory as the next chunk is processed. 
However, as diagnosed in \sect{moti}, this results in an abundant amount of non-coalesced access, becoming a reason for inefficiency.

To address this problem, we used \emph{lazy spilling} implemented using double-buffered shared memory for each warp to reduce the amount of global memory access. 
\fig{shm} (right) indicates the shared memory location where the threads write at each step. 
For each warp, a shared memory region is allocated with the size of $2\cdot dim(block)\cdot \#threads$. \JL{is it a cell? or 8 cells?}
As explained in \sect{parallel}, the first thread ($t0$) starts writing on the first location of the shared memory, which is passed to the second thread before the next step. 
In each step, the threads read data from the shared memory, process a block, and overwrite the bottom cells in the block to the shared memory.
Then, all threads shift one block location for the next step.

\begin{figure*}[t]
    \includegraphics[width=0.96\textwidth]{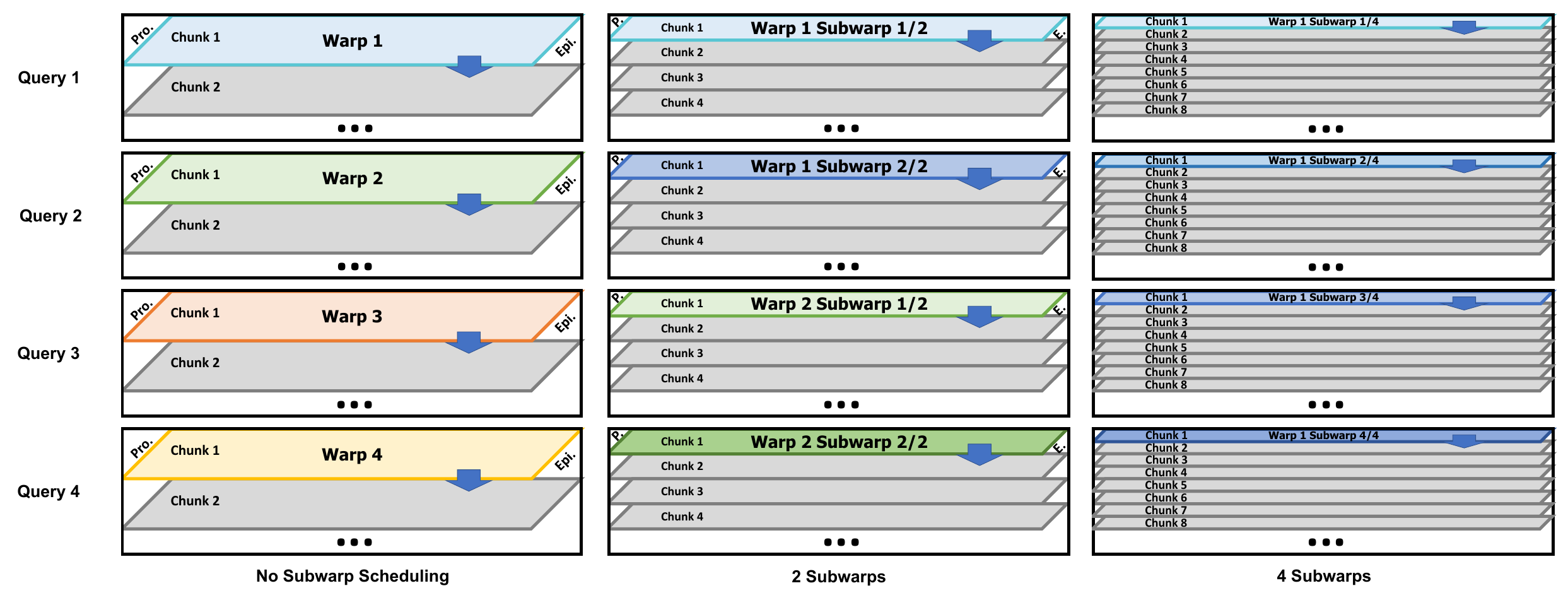}
    \caption{Subwarp scheduling.\vspace{-2mm}}
    \label{fig:vwc}
\end{figure*}

After the prologue, the threads leave a data trail, written by the last ($t31$) thread. 
Luckily, these are the data to be stored on the global memory. 
When the first thread ($t0$) reaches the end of the buffer at step 64, the shared memory has a consecutive region of data at the left side which are large enough to be processed by a warp. 
The threads stop processing and gather together to spill the data to the global memory in a coalesced write. 
Afterward, the threads wrap around in the shared memory, and the trail of block data starts forming on the right half of the shared memory, which is later spilled to the global memory.

The spilled data from the previous chunk must be read to process the next chunk  so that the first thread can use the data to calculate the cell values. 
This process is also performed in the same way using the double-buffered shared memory in the opposite direction of the writes.

\subsection{Subwarp Scheduling: Trade-off Between Parallelism and Workload Balancing}
The proposed strategy---a warp per query---can eliminate the workload imbalance between threads in a warp because they work together. 
However, as demonstrated in \sect{parallel}, a problem exists in the prologue and epilogue. 
The average thread utilization over these phases is 50\% (because it has a triangular form), and their size is proportional to the warp size. 
The DP table must be significantly larger than the number of threads in a warp to amortize this overhead. 
However, according to \fig{histo250}, the size of the table dimension is only around several hundred, which is only a few times larger than the size of 32, and sometimes even smaller.

We chose to split the warp into multiple \emph{subwarps}  and let them process a single query, as in \fig{vwc}, similar to the idea of VWC~\cite{vwc} used in graph processing. 
Its effect is essentially a trade-off between workload balancing and resource utilization.

When many subwarps are used (i.e., the subwarp size is smaller), the sizes of the prologue and epilogue become smaller; therefore, the average utilization increases. 
However, multiple subwarps make the scheme suffer more often from warp divergence due to workload imbalance. 
In contrast, when fewer subwarps are used (i.e., the subwarp size is larger), it is less likely to suffer from imbalance at the cost of lower utilization from increased portions of the prologue and epilogue.
Empirically, we found that using two or four subwarps yields the best results in our setting (see \sect{eval}).

One potential drawback is that using subwarps complicates the memory access optimization. 
If the same double-buffered technique is used, the number of threads accessing the consecutive memory address becomes less than 32.  
Subwarps larger than eight threads do not incur a considerable problem for architectures later than NVIDIA Volta
~\cite{khairy2018exploring}. 
For older architectures, the problem can be solved by allocating $N+32$ slots of shared memory instead of $2N$. 
By making the active region rotate around and making all threads in the entire warp 
write to 32 slots of data to the global memory together, a full coalescing can be achieved at the expense of a slightly higher shared memory capacity requirement.

\section{Evaluation}
\label{sec:eval}
\subsection{Experimental Setup}
\label{sec:env}
We evaluated \aligner on two platforms. 
First, as an `affordable’ system, we used a GTX1650 GPU card based on the Turing~\cite{turing} architecture. 
The server has a six-core Intel i5-9600K CPU with 16 GB of RAM. 
We also conducted experiments on a `high-end’ system with an RTX3090 GPU card based on the Ampere~\cite{ampere} architecture. 
The server has a single socket, 12-core AMD EPYC 7272 CPU and 128 GB of RAM. 
Both machines run on Ubuntu 18.04 with CUDA version 11.2 and NVIDIA driver 460.27.04.

\input{eval/kernel_all}

For comparison, we used the seed extension kernels from the following libraries as the baseline listed in \tab{baselines}. 
\textbf{SOAP3-dp~\cite{soap3dp}} is an early algorithm for GPU-assisted short read alignment, which utilizes inter-query parallelism. 
\textbf{CUSHAW} is a family of GPU-assisted short read alignment algorithms.
While most of its members \cite{cushaw1,cushaw2,cushaw3} use GPU only for seeding phase, \textbf{CUSHAW2-GPU}~\cite{cushaw2gpu} accelerates the seed extension part with a GPU using a similar strategy to SOAP3-dp. 
Better performance is obtained by compacting the global memory storage format and using CUDA texture memory.
\textbf{NVBIO~\cite{nvbio}} is a library of reusable components designed by NVIDIA to accelerate bioinformatics applications using CUDA. 
\textbf{SW\#~\cite{korpar2013sw}} is an algorithm that utilizes intra-query parallelism. 
It splits the DP table into multiple anti-diagonal partitions and processes one partition per each kernel launch. 
Because it launches the kernel multiple times per query, it targets very long sequences. 
\textbf{ADEPT~\cite{adept}} is the most recent work among algorithms that use intra-query parallelism. 
It uses shuffle instruction along with binary masking to make use of the anti-diagonal parallelism.
Lastly, \textbf{GASAL2~\cite{gasal2}} is the state-of-the-art library.  
It achieves superior performance by further executing the sequence packing on GPU devices.

\begin{table}[b]

\small
    \centering
    \caption{Baseline Kernels Under Comparison}
    \begin{tabular}{lccc}
    \toprule
        Kernel & Parallelism & Bitwidth & Mapping \\
        \midrule
        SOAP3-dp~\cite{soap3} & inter-query & 2 bits & one-to-one \\
        CUSHAW2-GPU~\cite{cushaw2gpu} & inter-query & 2 bits & one-to-many \\
        NVBIO~\cite{nvbio} & inter-query & 2,4,8 bits & one-to-many \\
        GASAL2~\cite{gasal2} & inter-query & 4 bits & one-to-one \\
        SW\#~\cite{korpar2013sw} & INTRA-query & 8 bits & one-to-many \\
        ADEPT~\cite{adept} & INTRA-query & 8 bits & one-to-one \\  
        \bottomrule
        \multicolumn{4}{r}{\scriptsize *(All kernels have been modified to have at least four-bit  } \\
        \multicolumn{4}{r}{\scriptsize GPU-assisted packing and to support one-to-one mapping.)}
    \end{tabular}
    \label{tab:baselines}
   
\end{table}

For a fair comparison, we have put our best efforts into optimizing the existing methods under the same environment.
Most importantly, we assume on-GPU sequence packing for all methods. 
GASAL2 is known to achieve state-of-the-art performance using a custom sequence packing kernel executed on GPUs.  
While the strategy is successful, the packing is orthogonal to the seed extension itself and the same method can be applied to all other kernels.  
We have modified SOAP3-dp, CUSHAW2-GPU, and ADEPT kernels to support five possible types of literals by unifying the kernels to work on four-bit packing.
We left SW\# to use its original eight-bit packing, because modifying it to support packing would 
complicate the memory access behavior.
In addition, we have modified some kernels (CUSHAW2-GPU, SW\#, and NVBIO) that only have one-to-many alignment modes to support one-to-one alignment.



\subsection{Kernel Performance Measurements} 
In this subsection, we compare the performance of the seed extension kernels without load imbalances using input reads of equal lengths.
To measure the performance under various input lengths, we used an in-house sequence read simulator similar to Wgsim~\cite{wgsim} to generate synthetic reads for each length in the range of 64 to 4096. 
Each kernel processed 5,000 reads per call 200 times, and these results were averaged to output the results in \fig{kernelall}. 
The execution times were measured with the $cudaEventElapsedTime()$ API.

\fig{kernelall} (a) and (b) present the performance comparison of the methods on the GTX1650 card, and (c) and (d) present that on the RTX3090 card. 
For lengths equal to or longer than 128 bp, \aligner outperforms all other methods. 
For a very short length of 64 bp, the execution time of NVBIO is slightly shorter than \aligner (0.42 ms vs 0.51 ms in GTX 1650 and 0.21 ms vs 0.24 ms in RTX 3090). 
This is reasonable because with intra-query parallelism, the effect of resource under-utilization from prologue and epilogue is relatively significant.
However, the break-even point is found at 128 bp, where the reduced global memory access of \aligner becomes dominant. 
The methods other than \aligner that utilize intra-query parallelism (SW\# and ADEPT) perform poorly compared to the others.
SW\# is especially slow because it divides a single DP table into multiple kernel calls where one kernel call processes an anti-diagonal group, resulting in very low resource utilization.

A rather surprising observation is that GASAL2 does not always show superior performance over other baselines. 
While GASAL2 reports multi-fold speedup compared to its previous work, its speedup is mainly from the on-GPU packing, not the extension strategy.
Because we combined CUSHAW2-GPU with the on-GPU packing module taken from GASAL2, it shows comparable results on GTX1650 for all lengths and slightly better performance on RTX3090 for long sequence lengths. 

The best speedup obtained at short lengths by \aligner is observed at 512 bp. 
The speedup is 27.7\% on GTX1650 and 43.6\% on RTX3090 against GASAL2. 
At longer sequence lengths, some baseline kernels fail to run due to structural limitation (ADEPT) or bounded device memory (NVBIO and SOAP3-dp). \JL{who fails for what}
The performance trend becomes more consistent for longer lengths because the portion of the overhead of global memory access in execution time diminishes. 
Another cause for this trend would be the decrease of the relative size of prologue/epilogue compared to the sequence length.
For inputs equal to or longer than 1,024 bp, the speedup of \aligner against GASAL2 is consistently around 30\% on GTX 1650 and 50\% on RTX 3090. 
Against CUSHAW2-GPU, the speedup is around 40\% on GTX1650 and 20\% on RTX 3090.

%


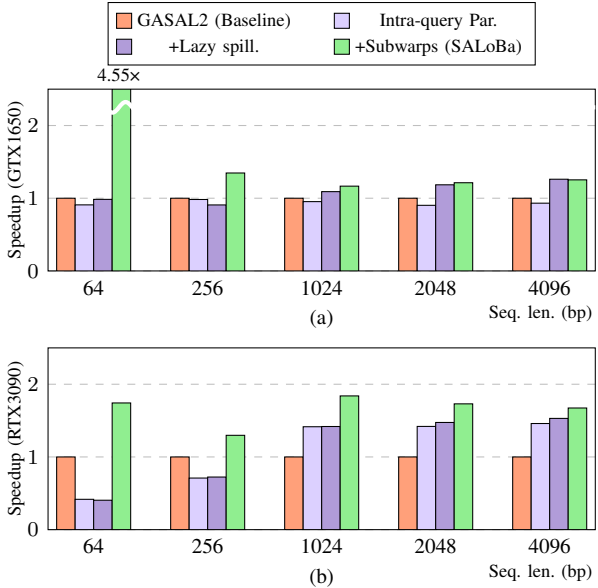
\begin{figure}[t]

\pgfplotstableread{
id	seq	gasal	Intra	memory	vwc
0	64	1	0.9085622052	0.9843882646	4.54658357
1	256	1	0.9826128764	0.9076944468	1.347011232
2	1024	1	0.9532737196	1.0905521	1.166026684
3	2048	1	0.9024406888	1.184560383	1.21306944
4	4096	1	0.9319012156	1.261305318	1.252616753
}\ablationdata

\begin{tikzpicture}
\pgfplotsset{
every axis title/.append style={at={(0.5,-0.45)},font=\footnotesize}
}
\pgfplotscreateplotcyclelist{custom}{
solid, thick, mark size=3pt, every mark/.append style={solid, thick, fill=color8}, mark=*\\%
dotted, thick, mark size=3pt, every mark/.append style={solid, thick, fill=color3}, mark=square*\\%
densely dotted, thick, mark size=3pt, every mark/.append style={solid, fill=color2}, mark=diamond*\\%
loosely dotted, thick, mark size=3pt, every mark/.append style={solid, fill=color5}, mark=triangle*\\%
solid, thick, mark size=3pt, every mark/.append style={solid, thick, fill=gray}, mark=*\\%
dashed, thick, every mark/.append style={solid, fill=color5},mark=otimes*\\%
loosely dashed, thick, every mark/.append style={solid, fill=gray},mark=*\\%
densely dashed, thick, every mark/.append style={solid, fill=gray},mark=square*\\%
dashdotted, thick, every mark/.append style={solid, fill=gray},mark=otimes*\\%
dashdotdotted, thick, every mark/.append style={solid},mark=star\\%
densely dashdotted,thick, every mark/.append style={solid, fill=gray},mark=diamond*\\%
}
\begin{axis}[
    title=(a),
	ybar=0pt, 
	width=\columnwidth, height=4cm,
	bar width=7pt, 
	ymin=0,
	ymax=2.5,
	xticklabels={64,256,1024,2048,4096},
	xtick=data, x tick label style={font=\footnotesize, yshift=1.5mm}, tick style = transparent,
	y tick label style={font=\footnotesize},
	ymajorgrids=true, major grid style={thin,dashed},
xlabel = {Seq. len. (bp)},
xlabel shift=-.3em,
xlabel near ticks,
xlabel style={at={(ticklabel cs:0.9)},font =\rmfamily\scriptsize},
    ylabel={Speedup (GTX1650)},
    ylabel style = {font=\scriptsize, yshift=-2mm},    
    restrict y to domain*=0:2.5, 
    visualization depends on=rawy\as\rawy,    
    after end axis/.code={ 
            \draw [ultra thick, white, decoration={snake, amplitude=2pt}, decorate] (rel axis cs:0.03,0.9) -- (rel axis cs:1,0.9);
        },
	legend style={fill=white, at={(0.5,1.13), font=\scriptsize},
	anchor=south,legend columns=2,
	/tikz/every even column/.append style={column sep=.5cm}},
    legend image code/.code={%
      \draw[#1] (0cm,-0.1cm) rectangle (0.2cm,0.1cm);
    }
	]  
	
\addplot[fill=color8] table [x=id, y=gasal] {\ablationdata};\addlegendentry{GASAL2 (Baseline)}
\addplot[fill=color3] table [x=id, y=Intra] {\ablationdata};     \addlegendentry{Intra-query Par.}
\addplot[fill=color4] table [x=id, y=memory] {\ablationdata};    \addlegendentry{+Lazy spill.}
\addplot[fill=color1] table [x=id, y=vwc] {\ablationdata};       \addlegendentry{+Subwarps (\aligner)}

\end{axis}
  \node at (rel axis cs:0.213,1.08) {\scriptsize4.55\texttimes};
\end{tikzpicture}

\pgfplotstableread{
id	seq	gasal	Intra	memory	vwc
0	64	1	0.4173283144	0.40473199	1.742569834
1	256	1	0.7096728217	0.7236259009	1.297514026
2	1024	1	1.415456099	1.418678715	1.839390299
3	2048	1	1.41986486	1.474302873	1.730230026
4   4096	1	1.459795431	1.530150091	1.673205062
}\ablationdata

\begin{tikzpicture}
\pgfplotsset{
every axis title/.append style={at={(0.5,-0.45)},font=\footnotesize}
}
\pgfplotscreateplotcyclelist{custom}{
solid, thick, mark size=3pt, every mark/.append style={solid, thick, fill=color8}, mark=*\\%
dotted, thick, mark size=3pt, every mark/.append style={solid, thick, fill=color3}, mark=square*\\%
densely dotted, thick, mark size=3pt, every mark/.append style={solid, fill=color2}, mark=diamond*\\%
loosely dotted, thick, mark size=3pt, every mark/.append style={solid, fill=color5}, mark=triangle*\\%
solid, thick, mark size=3pt, every mark/.append style={solid, thick, fill=gray}, mark=*\\%
dashed, thick, every mark/.append style={solid, fill=color5},mark=otimes*\\%
loosely dashed, thick, every mark/.append style={solid, fill=gray},mark=*\\%
densely dashed, thick, every mark/.append style={solid, fill=gray},mark=square*\\%
dashdotted, thick, every mark/.append style={solid, fill=gray},mark=otimes*\\%
dashdotdotted, thick, every mark/.append style={solid},mark=star\\%
densely dashdotted,thick, every mark/.append style={solid, fill=gray},mark=diamond*\\%
}
\begin{axis}[
    title=(b),
	ybar=0pt, 
	width=\columnwidth, height=4cm,
	bar width=7pt, 
	ymin=0,
	ymax=2.5,
	xticklabels={64,256,1024,2048,4096},
	xtick=data, x tick label style={font=\footnotesize, yshift=1.5mm}, tick style = transparent,
	y tick label style={font=\footnotesize},
	ymajorgrids=true, major grid style={thin,dashed},
xlabel = {Seq. len. (bp)},
xlabel shift=-.3em,
xlabel near ticks,
xlabel style={at={(ticklabel cs:0.9)},font =\rmfamily\scriptsize},
    ylabel={Speedup (RTX3090)},
    ylabel style = {font=\scriptsize, yshift=-2mm},
	legend style={fill=white, at={(0.5,1.05), font=\footnotesize},
	anchor=south,legend columns=2,
	/tikz/every even column/.append style={column sep=.5cm}},
    legend image code/.code={%
      \draw[#1] (0cm,-0.1cm) rectangle (0.2cm,0.1cm);
    }
	]  

\addplot[fill=color8] table [x=id, y=gasal] {\ablationdata};
\addplot[fill=color3] table [x=id, y=Intra] {\ablationdata};       
\addplot[fill=color4] table [x=id, y=memory] {\ablationdata};       
\addplot[fill=color1] table [x=id, y=vwc] {\ablationdata};       

\end{axis}
\end{tikzpicture}

 \caption{Ablation study for (a) GTX1650 and (b) RTX3090. } 
   \label{fig:ablation}
  \end{figure} 


\subsection{Ablation Study}

\fig{ablation} breaks down the contributions of the three techniques composing \aligner.
The speedups are normalized against GASAL2, which performs reasonably well along all sequence lengths we target.
The most effective technique for shorter lengths ($\leq$1024) is subwarp scheduling, which is expected because at shorter inputs, the portions of the prologue and the epilogue are very large. 
Therefore, switching to intra-query parallelism yields performance degradation. 
Using subwarp scheduling directly reduces the overhead and provides a substantial speedup.

A seemingly large speedup is observed at 64 bp for both GTX1650 and RTX3090, which is counter-intuitive because there is less intra-query parallelism that \aligner can exploit.
However, further analyzing \fig{kernelall} (a) and (c) reveals that it is the inefficiency of GASAL2, not speedup of \aligner. 
GASAL2 has a relatively large memory initialization cost at the beginning, and using a small sequence length fails to amortize it.
Compared to NVBIO instead of GASAL2 at 64 bp, it correctly reflects the intuition that inter-query parallelism is better suited for very short queries.

As the input becomes longer, the gain from using subwarps becomes marginal, and the effects of the other two techniques become significant. 
Intra-query parallelism and lazy spilling both contribute to less global memory access. 
The former reduces the amount of data stored in the global memory, whereas the latter reduces redundant access by coalescing access better. 
Therefore, these become the main driver for performance gains in longer input ranges. 
Interestingly, in  RTX3090, intra-query parallelism has more effect, whereas lazy spilling is more effective for GTX1650. 
We believe this is explainable by the fact that RTX3090 has a higher computation/memory bandwidth ratio. 
RTX3090 has a peak performance of 35.58 TFlops, and its memory bandwidth is 936.2 GB/s using GDDR6X. 
On the other hand, GTX has a peak performance of 2.98 TFlops with 128.1 GB/s memory bandwidth. 
The ratio is 38.91 Flops/B in RTX3090 and 23.82 Flops/B in GTX1650. 
This means that  RTX3090 is generally more bottlenecked at its memory. 
The major drawback of intra-query parallelism is the low CUDA thread utilization.
However, as RTX3090 is more bottlenecked in memory bandwidth, it is likely that there are more computational resources to process the data even when some are being idle.
This partially offsets the utilization problem in RTX3090.
\JL{check this again}

\subsection{Real-world Data Experiments}
To test \aligner under workload imbalance with the realistic distribution of workload sizes, 
we generate seeds from BWA-MEM~\cite{bwamem} using real-world datasets.
It takes a whole reference genome sequence and many 
sequence reads to generate seeds that are composed of multiple pairs that can be processed by extension kernels.

For the reference genome sequence, we used the latest release of the human genome assembly project, Build 38 patch release 13 (GRCh38.p13)~\cite{refsequence} that has 3.1G base pairs (bp).  \JL{SY: how long is it?}
For the input sequence reads, we used two datasets downloaded from the sequence read archive~\cite{sequencearxiv}.
The \textbf{dataset A} (SRR835433) is from a 2nd generation sequencer Illumina MiSeq and represents short reads 
where each sequence has a length of 250 bp. 
The dataset comprises 8.3M sequences which are randomly read genome parts from a human. 
Each sequence is read twice, resulting in the total number of base pairs to be 250$\times$8.3M$\times$2=4.1 Gbp.
The \textbf{dataset B} (SRP091981) is from a 3rd generation sequencer PacBio RS and represents long reads.
It also contains randomly read genome parts from a human where there are 82K sequences with variable lengths that averages around 2,000 bp to a total of 182.4 Mbp.
The distributions of the seeds processed by BWA-MEM~\cite{bwamem} are presented in \fig{histo250}. 

\input{eval/realkernel}

The results are plotted in \fig{realkernel}.
\fig{realkernel} (a) shows the performance for short read dataset A.
The best speedup of \aligner over the baseline GASAL2 is 32.5\% 
for GTX1650 and 20.2\% for RTX3090.
SOAP3-dp could not complete the workload on GTX1650 as some of the inputs exceeded the length it could process.
The speedup values observed from \aligner are slightly larger than that of \fig{kernelall} (a) and (c) due to the fact that \aligner suffers less from workload imbalance.
While the performance of SOAP3-dp and NVBIO are inferior to GASAL2, CUSHAW2-GPU exhibits some speedup over GASAL2 (with the help of GASAL2's on-GPU packing kernel).
However, its speedup is smaller than that of \aligner.
ADEPT achieves a similar speedup compared to \aligner only on RTX3090, but the speedup comes from placing all the intermediate values in the shared memory (no global memory access), which fundamentally limits ADEPT up to 1024 bp.

An interesting difference between GTX1650 and RTX3090 is that the optimal subwarp size was 16 for GTX1650 and eight for RTX3090 (see \fig{realkernel} (c)). 
This outcome is due to the different effectiveness of each technique.
According to \fig{ablation}, the benefit of applying subwarp scheduling for shorter sequences was 2.26$\times$ for GTX1650 and 2.85$\times$ for RTX3090 in geomean.
In GTX1650, the benefit of using subwarps is partially offset by the overhead from workload imbalance, shifting the optimal configuration to 16 threads per subwarp. 

\fig{realkernel} (b) reveals the performance for dataset B with longer reads.
On this dataset, SOAP3-dp, ADEPT, and NVBIO fail to run due to their input length limitations.
The speedup of \aligner greatly improves compared to that of \fig{kernelall} (b) and (d) because the amount of workload imbalance increases which works in favor of \aligner.
The best speedup is around 2.1$\times$ for both GTX1650 and RTX3090.
Subwarps with 16 threads per subwarps gained the best performance, which is the sweet spot between exploiting intra-query parallelism and reducing the workload imbalance.

\section{Related Work}
\subsection{GPU-based Smith--Waterman Algorithm Implementations}

 Accelerating Smith--Waterman algorithm~\cite{sw} with GPUs has been studied for many years.
Some earlier work based on OpenGL library~\cite{liu2006gpu, liu2006bio} addressed issues for mapping the algorithm on graphics pipeline. 
A popular CUDA implementation can be found from the CUDASW++ family~\cite{cudasw1,cudasw2,cudasw3}.
They suggest utilizing both the intra- and inter-query parallelism based on a pre-determined sequence length threshold~\cite{cudasw1}.
Similar approaches can be found from GSWABE~\cite{gswabe} and gpu-pairalign~\cite{gpu-pairalign}.
In addition, focused on all-to-all patterns for protein DB alignments, query profiling optimization~\cite{cudasw2} and CPU-GPU hybrid parallelism~\cite{cudasw3} have been suggested.
However, such optimizations are often too specific to protein alignments, and cannot be easily generalized to other domains such as DNAs.

In such regard, the CUDAlign family~\cite{cudalign1, cudalign2, cudalign3, cudalign4} focus on a general Smith--Waterman algorithm using intra-query parallelism.
CUDAlign~\cite{cudalign1} splits the DP table into multiple blocks and distributes them to threadblocks.
Then, the communication between the threadblocks are done using a dedicated region in the global memory called the `horizontal bus' and `vertical bus'.
The algorithm is later extended to support linear space algorithm to reduce the memory size~\cite{cudalign2}.
Its recent versions support multi-GPU alignment with execution time prediction~\cite{cudalign3} and traceback~\cite{cudalign4}.
When the DP table is too large, approaches such as MultiBP~\cite{multiBP}, MASA~\cite{masa}, and SW\#~\cite{korpar2013sw} suggest block pruning~\cite{cudalign21} that removes some blocks that can never achieve the optimal score. 

However, these algorithms are optimized for huge sequences that sometimes do not even fit into a single GPU card~\cite{khajeh2010acceleration}.
For DNA alignment softwares with seed-and-extend strategies, the input sequence length for the extension ranges around several hundreds even with the long DNA reads.
Because of this, libraries that rely on inter-query parallelism~\cite{nvbio,gasal2} often perform much better for DNA alignments as demonstrated in \cite{adept}.
On the other hand, \aligner outperforms the previous approaches on DNA alignment scenarios by adopting several careful optimizations.

\subsection{GPU-accelerated Sequence Alignment Softwares}
Some approaches design an end-to-end DNA alignment software that are friendly to GPU architectures.
SARUMAN~\cite{saruman} and GPU-RMAP~\cite{gpu-rmap} are some early methods that
implement hashtable lookups on GPUs to perform short read mappings.
With the increased use of the Burrows--Wheeler transform (BWT)~\cite{bwt} based on suffix-trees, BarraCUDA~\cite{barracuda}, SOAP3~\cite{soap3} and CUSHAW1~\cite{cushaw1} were introduced with GPU implementations of BWT.
SOAP3-dp~\cite{soap3dp}, CUSHAW2-GPU~\cite{cushaw2gpu}, and LOGAN~\cite{zeni2020logan} are later methods that adopt seed-and-extend strategy, 
and Arioc~\cite{arioc} further expands the strategy to multiple GPUs.
However, these methods often sacrifice alignment quality for the sake of throughput.
For example, SOAP3 and LOGAN only supports limited number of mismatches, and CUSHAW2-GPU packs the sequences to two bits by converting the fifth `N' bases to random bases.

Although the approaches above show good throughput with only a small amount of quality degradation, 
the industry has grown to value quality over speed.
As high-quality algorithms such as BWA-MEM~\cite{bwamem} became a de facto standard, the usability of the GPU-aware alignment softwares were limited.
Some approaches~\cite{zaid1,zaid2,zaid3,adept} tackle this problem and design a seed extension kernel general enough to be used for BWA-MEM using intra-query performance. 
However, later approaches based on inter-query parallelism outperformed these kernels, which is the strategy adopted by the current state-of-the-art methods such as NVBIO~\cite{nvbio} or GASAL2~\cite{gasal2}.

Finally, it is worth mentioning cuBLASTP~\cite{cublastp}, a GPU accelerated protein search algorithm that uses a variant of seed-and-extend strategy.
However, cuBLASTP only accelerates the seeding part using GPU, and leaves the extension part to CPU for utilizing CPU-GPU hybrid parallelism.

\subsection{FPGA/ASIC Acceleration}
As interest in the genomics pipeline has increased, more researchers have considered designing dedicated accelerators using FPGAs or ASICs.
\cite{zhang2007implementation} provided a implementation of the Smith--Waterman algorithm on an Altera FPGA, followed by a few other studies using banded algorithms~\cite{chen2013hardware, blastp} or flexible systolic arrays~\cite{houtgast2015fpga,ahmed2015heterogeneous}. 
Moreover, DRAGEN~\cite{dragen} is an FPGA-based platform from Illumina that implements a full end-to-end analysis into an FPGA.
 Darwin~\cite{turakhia2018darwin} is an ASIC accelerator that speeds up the whole genome sequencing through the co-design of both the software and hardware, powered by gapped filtering. 
GenAx~\cite{fujiki2018genax} is an automata-based accelerator for both seeding and extension. 

Although these approaches provide significant speedup,
FPGAs or ASICs are much harder to design and it requires a long time to reach the market. 
Compared to them, GPU-based alignment software could be a quick solution easily adopted by any system that has GPUs attached.






\section{Discussion}

\subsection{CUDA Shuffle Instructions}
The CUDA shuffle instruction is an alternative to 
shared memory by allowing a direct register-to-register data exchange for inter--thread communication. 
However, using shuffle-based communication did not add any additional speedup on top of \aligner. 
This outcome aligns with the CUDA specification that the throughput of the shuffle instructions is almost the same as the nonconflicting shared memory reads~\cite{warpperf}.

One potential gain from shuffle instructions is reducing shared memory usage.  
However, the portion of data cached in the shared memory was negligible in the current scenario and did not provide noticeable speedup.

\subsection{Banded Algorithms}
Many seed extension methods use the banded algorithm to reduce the computational burden. 
Taking advantage of the fact that the best matching path of the DP table is usually near the diagonal, processing cells within some predetermined width (the \emph{band}) often yields solutions of sufficient quality. 
However, most GPU-based extension libraries do not exploit this.

One problem is that the band sizes for each query are often different, which worsens load balancing. 
However, we envision that banded algorithms would have more benefits and could be adopted for better performance with longer reads. 

\subsection{Multiple GPUs}
It is often beneficial to use multiple GPUs, especially when equipped on a single machine. 
We believe that extending this work by splitting the queries into equal numbers and assigning them to multiple GPUs would be straightforward.  
One possible drawback of such a strategy would be the load imbalance between the GPUs. 
However, the penalty would be small compared to the thread-level imbalance problem. 
If this becomes significant, it could be solved by dynamic assignment or preprocessing with approximate sorting. 

\section{Conclusion}
We proposed \aligner, a new GPU-based seed extension library, exhibiting  state-of-the-art performance over the existing work. The experiments reveal that the software performs well on modern devices, and the performance gain increases with recent GPUs. We believe this software will be useful in fields where sequence alignment times are critical for diagnosing fatal diseases.

\section*{Acknowledgements}
\small{
This work has been supported by the National Research Foundation of Korea (NRF) grant funded by the Korea government (MSIT) (2022R1C1C1008131,  
2022R1C1C1011307)  
and 
Institute of Information \& communications Technology Planning \& Evaluation
(IITP) grant funded by the Korea government (MSIT) 
(2020-0-01361,
Artificial Intelligence Graduate School Program (Yonsei University),  
2021-0-00853, Developing Software Platform for Programming of PIM
).
}

\JL{Fix the capitalizations in ref}

\bibliographystyle{IEEEtranS}
\balance
\bibliography{refs}

\end{document}